\documentclass[conference,compsoc]{IEEEtran}

\ifCLASSOPTIONcompsoc
  \usepackage[nocompress]{cite}
\else
  \usepackage{cite}
\fi

\ifCLASSINFOpdf
  \usepackage[pdftex]{graphicx}
\else
\fi

\usepackage{amsmath}
\usepackage{dblfloatfix}
\begin{document}

\title{IEEE BigData 2021 Cup: Soft Sensing at Scale}

\makeatletter
\newcommand{\linebreakand}{%
  \end{@IEEEauthorhalign}
  \hfill\mbox{}\par
  \mbox{}\hfill\begin{@IEEEauthorhalign}
}
\makeatother
\author{
  \IEEEauthorblockN{Sergei Petrov}
  \IEEEauthorblockA{\textit{Seagate Technology, MN, US} \\
    \textit{Stanford University, CA, US}\\
    sergei.petrov@seagate.com}
  \and
  \IEEEauthorblockN{Chao Zhang}
  \IEEEauthorblockA{\textit{Seagate Technology, MN, US} \\
    \textit{University of Chicago, IL, US}\\
    chao.1.zhang@seagate.com}
  \and
  \IEEEauthorblockN{Jaswanth Yella}
  \IEEEauthorblockA{\textit{Seagate Technology, MN, US} \\
    \textit{University of Cincinnati, OH, US}\\
     jaswanth.k.yella@seagate.com}
  \linebreakand
  \IEEEauthorblockN{Yu Huang}
  \IEEEauthorblockA{\textit{Seagate Technology, MN, US} \\
    \textit{Florida Atlantic University, FL, US}\\
    yu.1.huang@seagate.com}
  \and 
  \IEEEauthorblockN{Xiaoye Qian}
  \IEEEauthorblockA{\textit{Seagate Technology, MN, US} \\
    \textit{Case Western Reserve University, OH, US}\\
    xiaoye.qian@seagate.com}
  \and
  \IEEEauthorblockN{Sthitie Bom}
  \IEEEauthorblockA{\textit{Seagate Technology, MN, US} \\
    sthitie.e.bom@seagate.com}
}

\maketitle

\begin{abstract}
“IEEE BigData 2021 Cup: Soft Sensing at Scale” is a data mining competition organized by Seagate Technology, in association with the IEEE BigData 2021 conference. The scope of this challenge is to tackle the task of classifying soft sensing data with machine learning techniques. In this paper we go into the details of the challenge and describe the data set provided to participants. We define the metrics of interest, baseline models, and describe approaches we found meaningful which may be a good starting point for further analysis. We discuss the results obtained with our approaches and give insights on what potential challenges participants may run into.\\
Students, researchers, and anyone interested in working on a major industrial problem are welcome to participate in the challenge!
\end{abstract}

\IEEEpeerreviewmaketitle

\section{Introduction}
\noindent
Over the last few decades, different industries have seen a dramatic increase in utilization of smart sensors in an attempt to control and better understand various production processes. Data coming from smart sensors can provide detailed information on the process under control as well as indicate issues that could have occurred. The term "soft sensing" is in general used for the approaches that are  utilized to estimate certain physical quantities or a product quality  in  the  industrial  processes. \\\\
In the wafer factories at Seagate Technology, sensors are used to monitor the process stability of wafers, therefore the sensor data may potentially contain all the necessary information regarding the state and quality of wafers. Wafer manufacturing entails a complex combination of multi-staged sequences of deposition, etch, and lithography processes with dedicated tools performing measurements at the end of each processing stage to assess product quality. The current quality control process involves engineers manually labeling wafers as passed or failed based on the data coming from the tools measuring physical properties of wafers. Such tools are solely used for wafers quality control purposes, which makes the entire manufacturing process even more costly. The amount of data as well as absence of a straightforward standard procedure make manual labeling a very time consuming and ambiguous process. Thus, it is natural to look for ways to make it more robust and eventually to automate it.\\\\
However, the sensor data is high-dimensional, sometimes redundant, prone to outliers, and are not interpretable by humans. Due to these difficulties, such kind of data can be challenging to process not only for humans but for machine learning algorithms as well. Another challenge is the highly imbalanced data set with failed examples being severely underrepresented.

\section{Data Acquisition}
\noindent
The wafer manufacturing process is complex and expensive. During production a wafer goes through multiple processing stages which include deposition, coating, lithography etching, and polishing. Different types of wafers go through different manufacturing stages. After each processing stage, physical properties of the wafer are measured with metrology tools for quality control purposes. Each metrology step usually has several different measurements, and the same measurements may be performed after different stages. A very generalized production scheme is shown in Fig. \ref{wafer_proc}.\\

\begin{figure*}[!t]
\centering
\includegraphics[width=\textwidth]{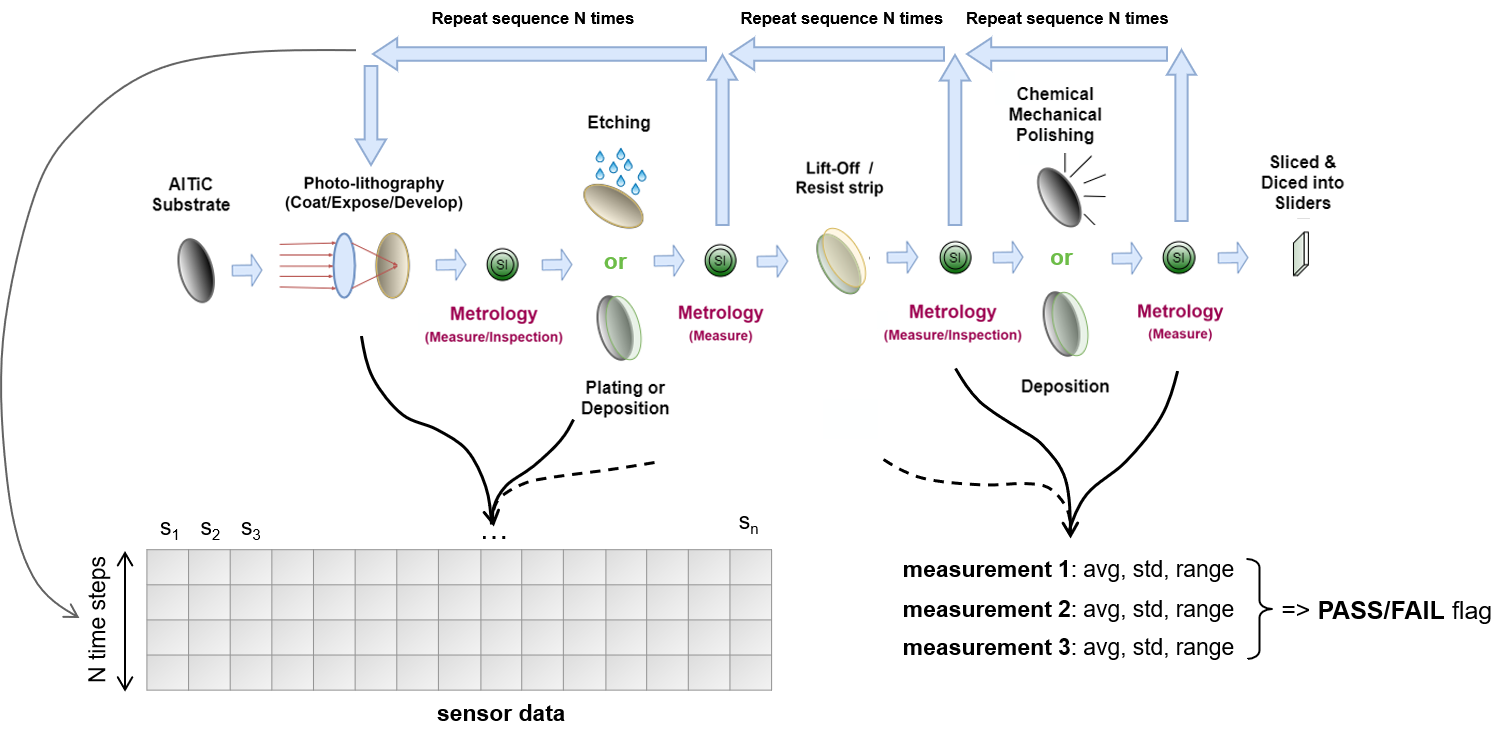}
\caption{A simplified wafer manufacturing process}
\label{wafer_proc}
\end{figure*}
\noindent
Metrology is an extremely important stage that captures necessary information to make informed decisions regarding wafer process quality. However, it comes at high capital costs and considerably increases the production time. Apart from the time it takes to make the measurements themselves, it requires a lot of engineering time to set up correct workflows, decide on what measurements to run on each stage and then to process the output data. In each production stage, the state of the wafer is recorded by dozens to hundreds of sensors placed in the processing machines. These sensors collect information every few seconds, and all the data obtained by sensors and metrology tools are stored in the database.\\\\
Since several measurements can be made at each manufacturing stage, and the same measurements can be applied at different stages, a single time series of sensor data maps to several non-unique measurements. Results of each measurement are summarized in a few statistics (average, standard deviation etc.), and these numbers are used by engineers to label as pass or fail.\\\\
In the raw data, label is assigned on the measurement step level which comprises several distinct measurements. If at least one critical measurement within a step was marked as failed at the quality control stage, the entire step gets labeled as failed. As a single time series can map to more than one step and different series map to different sets of steps, the classification task becomes a multi-task classification problem.\\\\
The sensor data used for this challenge was recorded over 92 weeks. Though sensors capture information every few seconds, there is a lot of redundancy due to high correlation between adjacent measurements. Because of this the data is aggregated into relatively short sequences, which also makes the data set significantly more manageable. The missing data was imputed within stages by forward filling  and  backward  filling from the neighboring values. If there were no values to rely on in a particular column, the mode of all the data was used. We provide a detailed description of the data in the next section.
\section{Outline of the IEEE BigData 2021 Cup}
\subsection{Data provided to participants}
\noindent
The goal of this challenge is to develop a solution in a form of a machine learning model to go directly from sensor data and associated categorical variables to labeling a wafer as passed or failed. Such a solution can potentially help to skip some measurement steps reducing the overall time and cost of wafer manufacturing.
Thus, participants are expected to train a model performing binary classification using time series multivariate data as an input. \\\\
The data set is provided by Seagate and collected from factories in both Minnesota and Ireland. It contains high dimensional time-series sensor data along with relevant categorical variables. In the raw data, a unique identifier used to aggregate data into time series is a combination of the wafer identifier and a processing stage identifier. The same combination of variables (further referred to as the key) is used to map the resulting time series data to the labels stored in a separate table along with measurement results. Preprocessing of raw data included the following steps: 
\begin{itemize}
\item Redundant, erroneous and otherwise unhelpful data was dropped: columns containing a single constant value; measurement steps having a single measurement; duplicate rows; columns in the validation/test set not having counterparts in the training set
\item The data were scaled from 0 to 1 using ranges from the columns in the training set.
\item Data were aggregated to sequences by the key (transformation from 2D to 3D)
\item All sequences were padded to the same length.
\end{itemize}
The distribution of time series lengths before padding is shown in Fig.\ref{fig_seq_length}. As we can see, the majority of series have only two steps in them, and two is the maximum series length. Series are short, which may or may not restrict the performance of some algorithms specifically designed to work with time series data.\\\\
\begin{figure}[!t]
\centering
\includegraphics[width=0.9\linewidth]{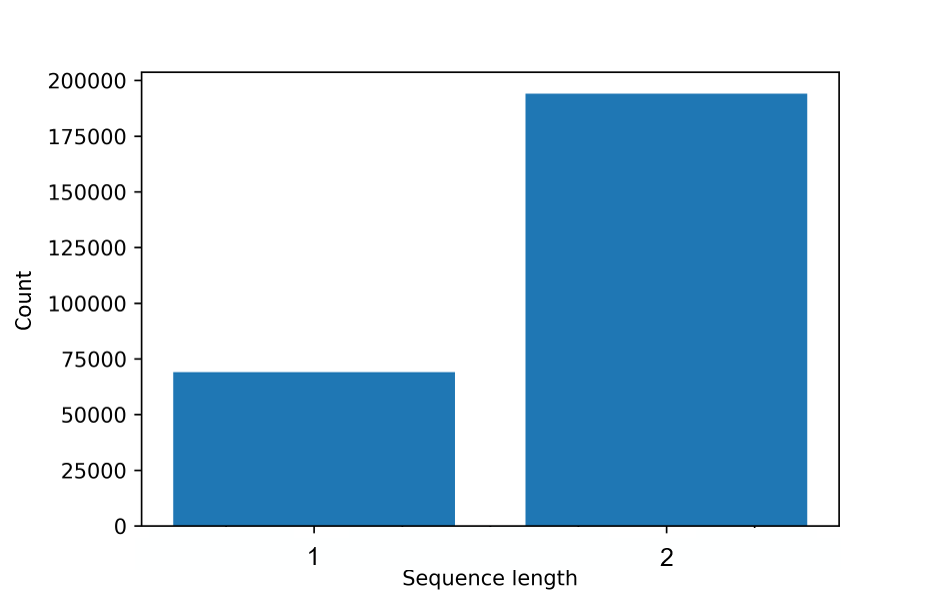}
\caption{The distribution of time series lengths before padding.}
\label{fig_seq_length}
\end{figure}
\noindent
Comparison of distributions of data coming from different classes can be a good way to explore the difference between classes. In this case, however, there is no simple threshold value that can be used to separate failed examples from passed ones. Even if we examine the data on an individual sensors level, passed and failed samples still cannot be distinguished by comparing their histograms. In Fig. \ref{fig_pass_vs_fail_sensors} the distributions of scaled values per class are shown for six sensors randomly drawn from the data set. The difference in distributions is caused by failed examples being underrepresented.
\begin{figure}[!t]
\begin{tabular}{ll}
  \includegraphics[width=0.45\linewidth]{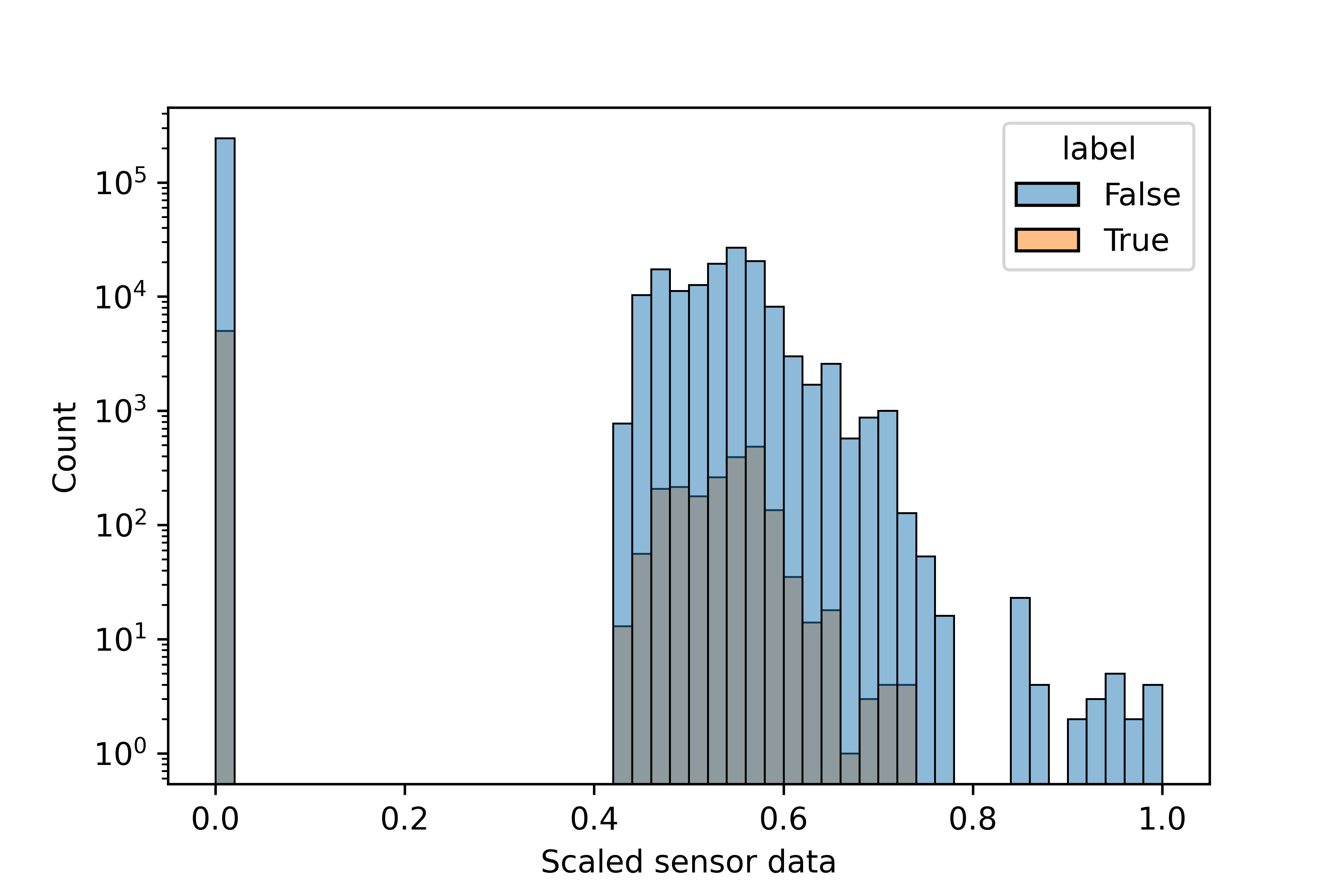}&
  \includegraphics[width=0.45\linewidth]{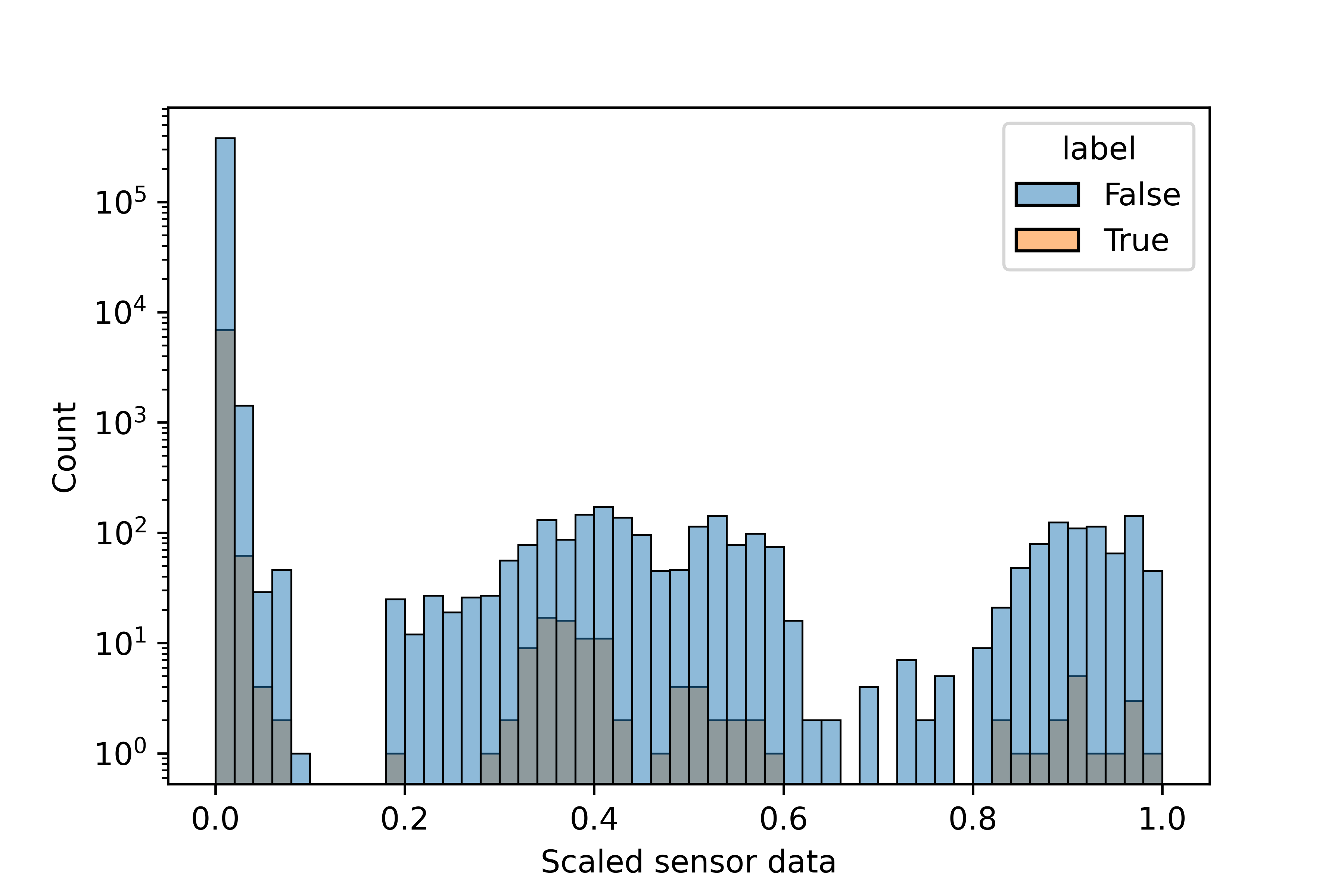} \\
  \includegraphics[width=0.45\linewidth]{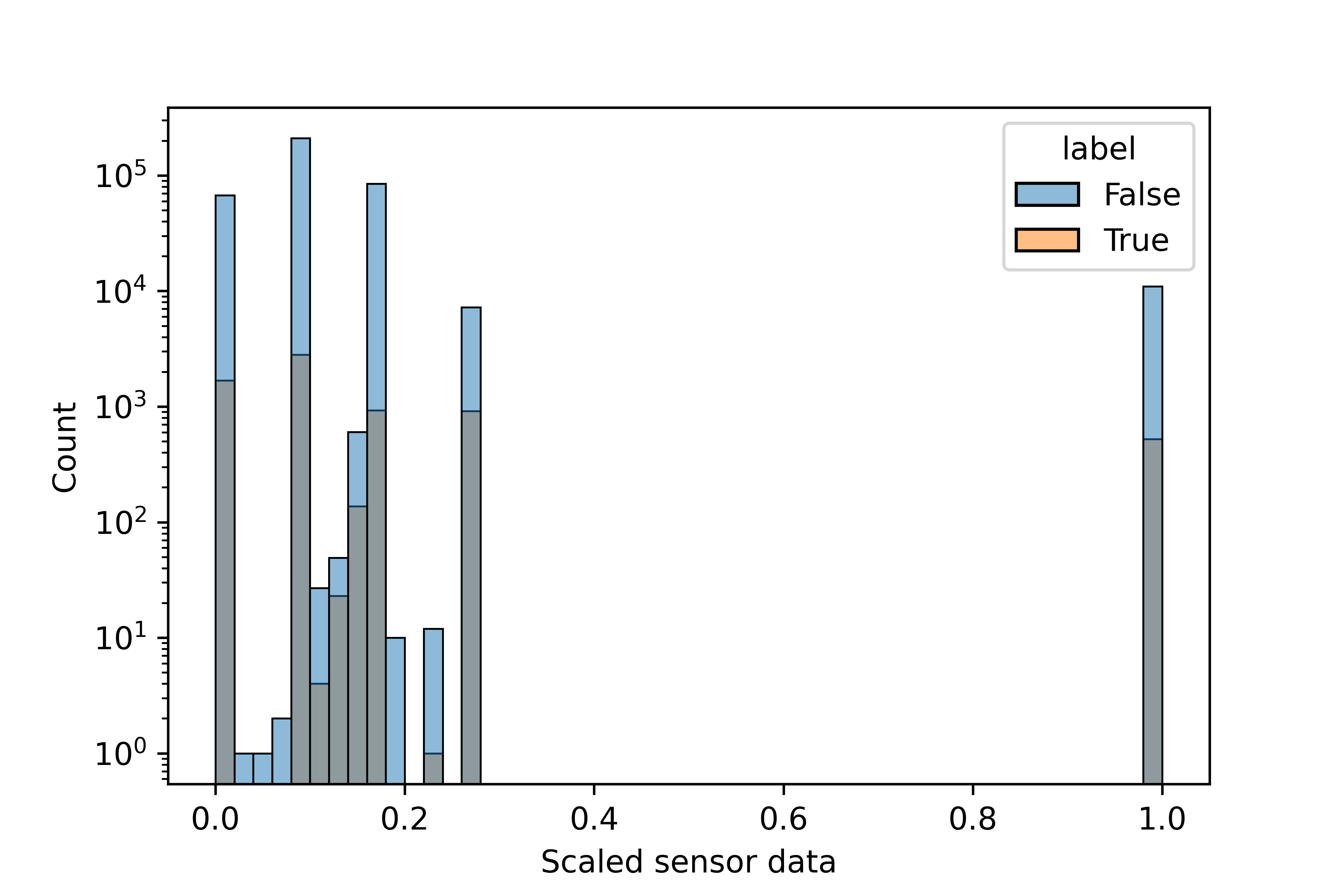}&
  \includegraphics[width=0.45\linewidth]{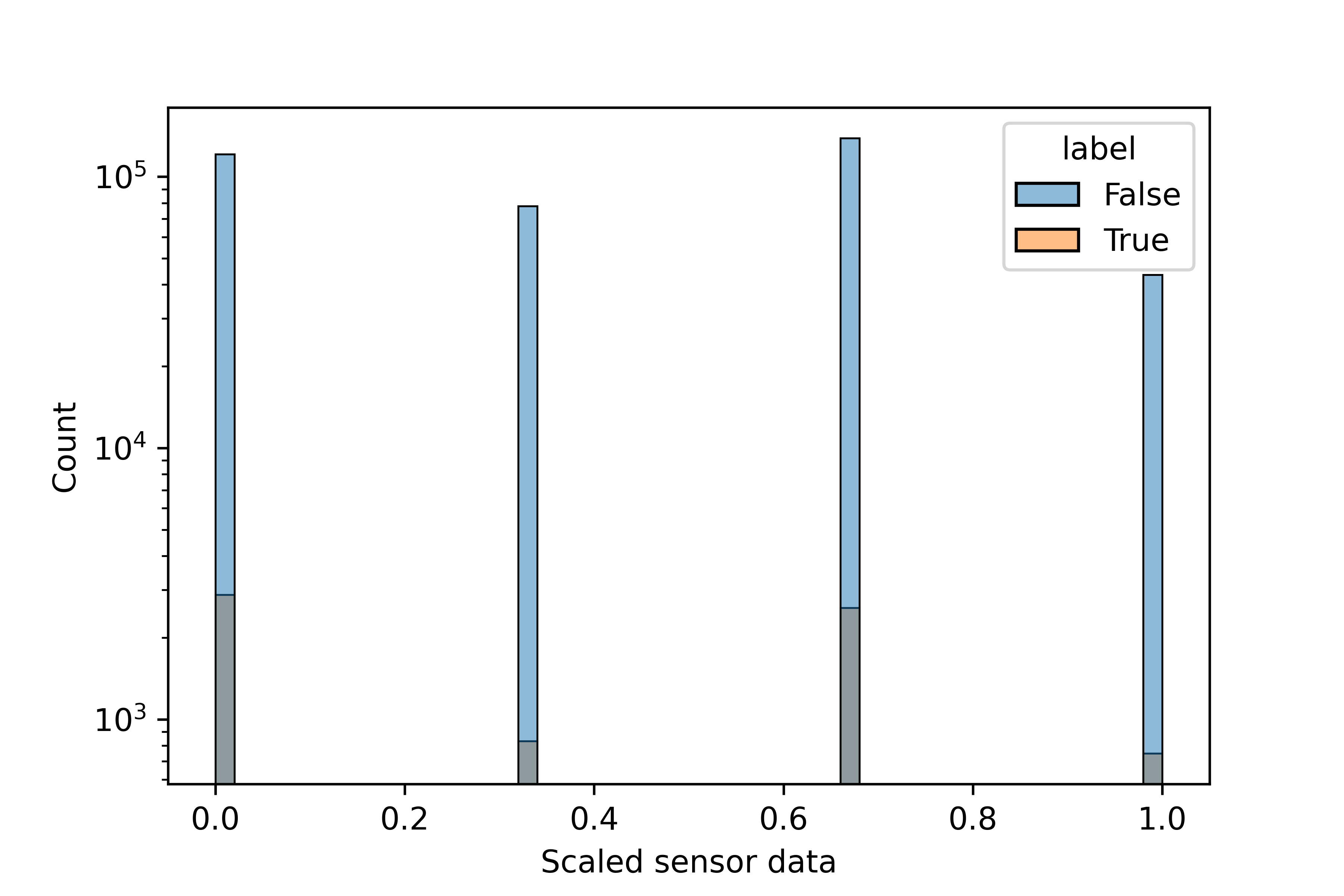} \\
  \includegraphics[width=0.45\linewidth]{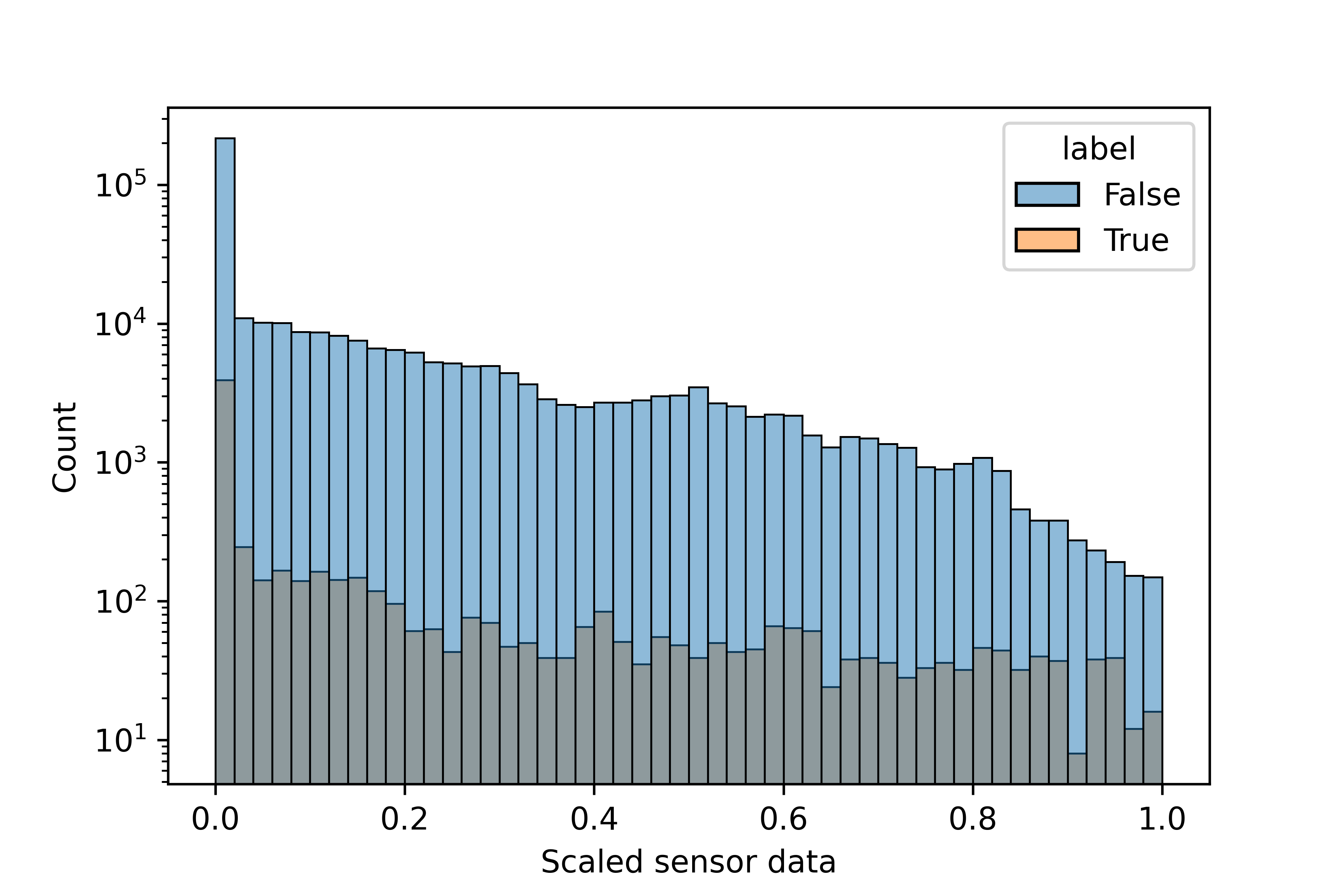}&
  \includegraphics[width=0.45\linewidth]{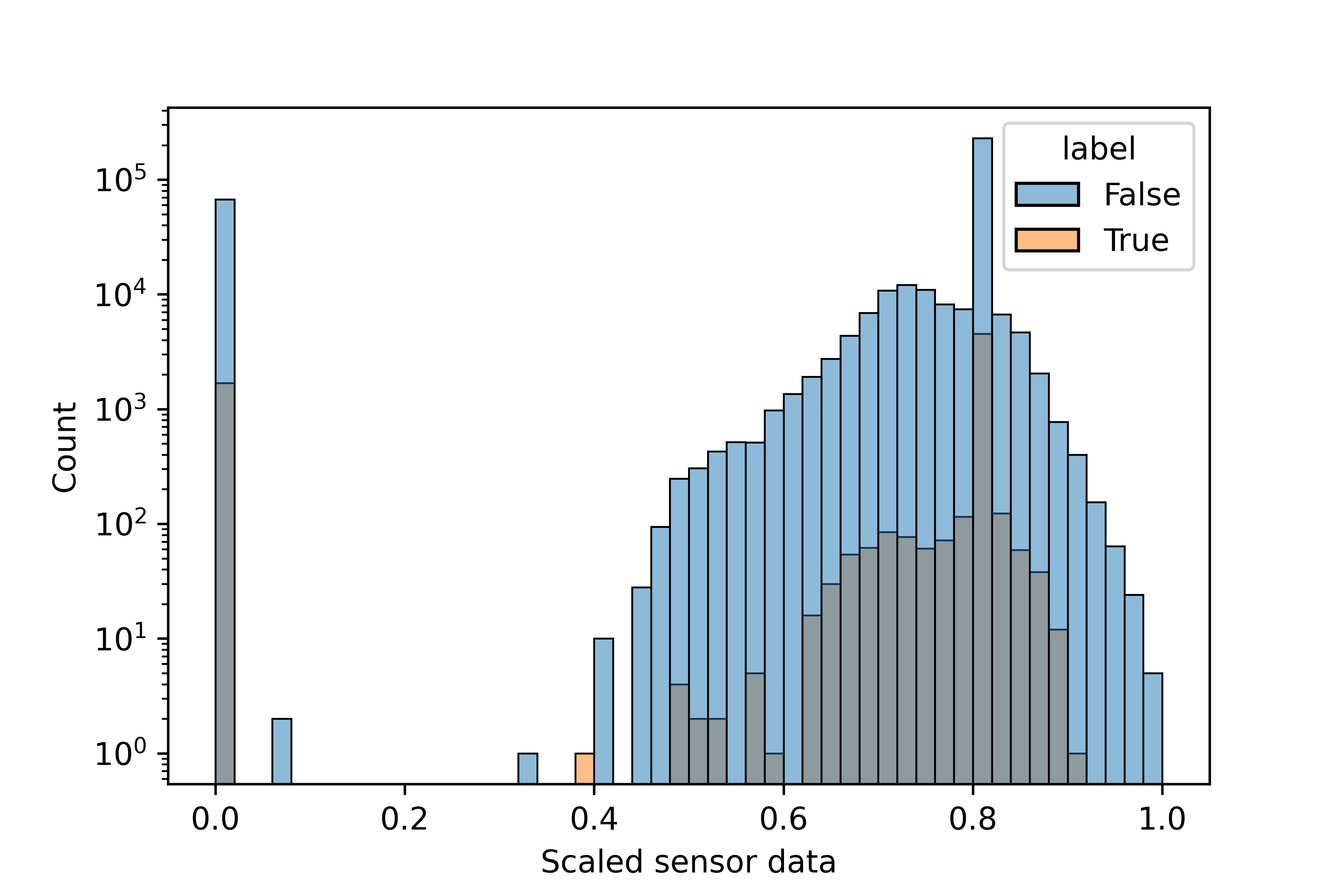} \\
\end{tabular}
\caption{The distributions of scaled sensor data per class per individual sensor. Six sensors were randomly drawn from the data set.}
\label{fig_pass_vs_fail_sensors}
\end{figure}
\noindent
This highlights again, that the task of soft-sensing data classification is not straightforward and requires applying advanced techniques. Another important aspect that is necessary to take into account is the fact that the data are highly imbalanced with the fraction of positive examples being in the range of 1 to 2 percent.\\\\
The entire data set provided to participants is split into two separate tables. One contains the data recorded by sensors and a few categorical variables complementing it, another table contains labels and keys mapping them to the input data.\\\\
The input sensor data is time series data and has 3 dimensions: samples, time-steps, features. The number of time steps varies across samples in the raw data, so the data were padded to make them uniform. There are two kinds of features, float and integer, and the last column is a padding indicator - if the value is 1, the time-step is a padding step. Float features are the sensor data scaled from 0 to 1 (columns 727-815), and integer features are one-hot encoded categorical features (column 0-726). In the raw data there were 6 categorical variables, which were independently one-hot encoded, concatenated, and columns were shuffled. Thus, there are 6 positive indicator values within the first 727 columns for each time step, unless it's a padding step. In the validation and test sets there may be fewer positive values because of the possible mismatches in categorical variables between the training/validation/test sets in the raw data. \\\\
The table with labels is a two-dimensional array of data with each row being a label for an input sample.  A single output label is represented by a vector of length 22, which corresponds to positive and negative outcomes for each of 11 possible tasks (steps). This is a multi-task classification problem meaning for a single input example there could be multiple positive labels in the output vector. Since a vector represents labels for multiple tasks, it has the following structure: [task-0-negative, task-0-positive, task-1-negative, task-1-positive...]. For each pair of  columns, the possible values are: positive (0,1), negative (1,0), and missing (0,0).\\\\
Train/validation/test split is time-based, which means the chronological order in which data samples were recorded was preserved. While the labels for the training and validation sets are available to participants, the test set labels are hidden. 
%\vfill
\subsection{Evaluation metrics and baseline models}
\noindent
Solutions will be evaluated on the ROC-AUC metric calculated using predictions obtained on the test set. To obtain the average ROC-AUC, the AUC for each task needs to be calculated using non-missing labels, and then a weighted average of all tasks should be taken.\\\\
An example of calculating the AUC:\\
Assuming we have the following labels: [[1,0,0,0],[1,0,0,1],[0,1,0,0]], there are two tasks as there are two pairs of columns. There are 3 samples for the first task - 1 positive (0,1) and 2 negative (1,0), and the second task has one positive sample. The AUC for each of the tasks can be calculated independently with their corresponding samples. Then the average AUC can be obtained as (3*AUC1 + AUC2)/4.\\\\
There were several baseline models we have experimented.\\\\ Bidirectional LSTM model takes into account both left and right context by traversing the input sequence twice from left to right and from right to left \cite{Bi-LSTM}. The architecture was chosen due to the sequential nature of the data, as recurrent architectures are designed to work with time distributed data.\\\\
Variance weighted multi-headed quality driven autoencoder (VWMHQAE) was developed to tackle the same task of soft sensing data classification \cite{zhang2021autoencoder}. The architecture is a stacked autoencoder with multiple heads designed to perform binary classification for multiple tasks and with variance-based weights applied to the tasks. The architecture proved to be efficient on the soft sensing data classification task thus serves as a meaningful baseline model. It doesn't have capabilities to work with time series, so the input had to be flattened to make the model applicable.\\\\
Soft Sensing Transformer model utilized the time-series information and formatted the data into the same shape as embedded sentences, and applied a multi-task learning for this problem. The time-series input data is first fed into a self-attention layer, and applied with a global average pooling so that it can be connected to fully connected layers. The final layers of the model is a set of dense layers, and output layer is of size $n\_task * 2$, each pair of the output stands for a task. Within each pair, the larger value is taken as the classification prediction. For example, in case of two tasks, a prediction of [0.1, 0.2, 0.8, 0.7] is [positive, negative].

The results featuring the comparison with baseline models are shown in Fig. \ref{fig_auc_sst}. The average ROC-AUC score obtained with Soft Sensing Transformer is \textbf{0.685}, Bi-LSTM is \textbf{0.671}, and VWMHQAE is \textbf{0.670}.

\begin{figure}[!t]
\centering
\includegraphics[width=0.9\linewidth]{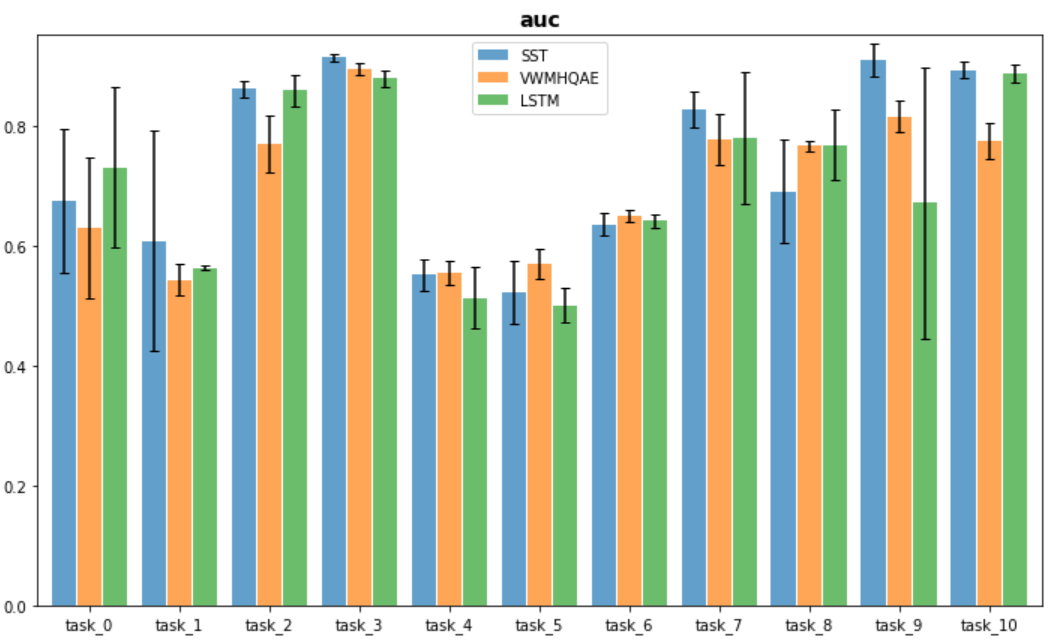}
\caption{AUC values obtained with Soft Sensing Transformer compared to the baseline models}
\label{fig_auc_sst}
\end{figure}

\section{Conclusions}
\noindent
In this work we described the scope of the BigData 2021 challenge. The goal of the challenge is to address the problem of soft-sensing data classification, which has the potential of making significant impact on the existing industrial workflows. We went into the details of the problem describing the associated processes and reasons behind why the problem exists.\\\\
The data and preliminary results are provided by Seagate Technology. The data provided to participants is the real-world soft-sensing data obtained with smart sensors during the manufacturing process and labeled by engineers, properly anonymized. We described the data in detail giving the overview of processing steps applied to the raw data set.\\\\
The description of the target metric was provided with some necessary explanations. We went over several machine learning models which could serve as baseline models to compare performance with.  These approaches can be used to borrow some ideas from, to gain better understanding of what challenges to expect when training other machine learning models.\\
Researchers, students, and anyone interested in tackling the challenge are welcome to participate.

\ifCLASSOPTIONcompsoc
  \section*{Acknowledgments}
\else
  \section*{Acknowledgment}
\fi
\noindent
The authors would like to thank Seagate Technology for the support on this study, the Seagate Lyve Cloud team for providing the data infrastructure, and the Seagate Open Source Program Office for open sourcing the data sets and the code.  Special thanks to the Seagate Data Analytics and Reporting Systems team for inspiring the discussions.

\bibliographystyle{IEEEtran}
\bibliography{ref}

\end{document}